\documentclass[12pt,preprint]{aastex}
\shorttitle{W49N OH Masers: anisotropic scattering and Zeeman pairs}
\shortauthors{Avinash A. Deshpande, W. M. Goss \& J. E. Mendoza-Torres}

\begin{document}

\title{OH Maser Sources in W49N: Probing Magnetic Field and Differential Anisotropic Scattering with Zeeman pairs using the VLBA}
\author{Avinash A. Deshpande\altaffilmark{1,2}, W. M. Goss\altaffilmark{3} and J. E. Mendoza-Torres\altaffilmark{4}}
\altaffiltext{1}{Raman Research Institute, Sadashivanagar, Bangalore, 560080 India; desh@rri.res.in} 
\altaffiltext{2}{University of Tasmania, Private Bag 21, Hobart 7001, TAS, Australia}
\altaffiltext{3}{National Radio Astronomy Observatory, P.O. Box O, Socorro, NM 87801 USA; mgoss@aoc.nrao.edu}
\altaffiltext{4}{Instituto Nacional de Astrof\'isica Optica y Electr\'onica, Tonantzintla, Puebla, 72840, M\'exico; mend@inaoep.mx}

\begin{abstract}
Our analysis of a VLBA 12-hour synthesis observation of the OH masers 
in a well-known star-forming region W49N has yielded valuable data 
that enables us to probe distributions of magnetic fields 
in both the maser columns and the intervening interstellar medium (ISM). 
The data consisting of detailed high angular-resolution images 
(with beam-width $\sim$20 milli-arc-seconds) 
of several dozen OH maser sources or {\it spots}, 
at 1612, 1665 and 1667 MHz, 
reveal anisotropic scatter broadening, with typical sizes of a few tens of 
milli-arc-seconds and axial ratios between 1.5 to 3. 
Such anisotropies have been reported earlier by 
Desai, Gwinn \& Diamond (1994) and interpreted as induced by the 
local magnetic field parallel to the Galactic plane.
However, we find a) the apparent angular sizes 
on the average a factor of about 2.5 less than those reported 
by Desai et al. (1994), indicating significantly 
less scattering than inferred earlier, 
and b) a significant deviation in the average orientation of the 
scatter-broadened images (by $\sim$10 degrees) from 
that implied by the magnetic field in the Galactic plane. 
More intriguingly, for a few Zeeman pairs in our set, 
significant differences (up to 6$\sigma$) 
are apparent in the scatter broadened 
images for the two hands of circular polarization, 
even when apparent velocity separation is less than 0.1 km s$^{-1}$. 
This may possibly be the first example of a Faraday rotation 
contribution to the diffractive effects in the ISM. 
Using the Zeeman pairs, we also
study the distribution of magnetic field in the W49N complex,
finding no significant trend in the spatial structure function.
In this paper, we present the details 
of our observations and analysis leading to these findings, 
discuss implications of our results
for the intervening anisotropic magneto-ionic medium, and  
suggest the possible implications for the structure of
magnetic fields within this star-forming region.
\end{abstract}

\keywords{masers --- ISM: molecules --- magnetic fields --- 
individual (W49N) --- radio lines: ISM --- structure function}

\maketitle

\section{Introduction}
W49N is a well-known and extensively-studied massive star-forming complex 
in the Milky Way. The OH and H$_2$O masers in this region represent some of
the most luminous such sources in our Galaxy. 
The W49N complex, in the 
Galactic direction $(\ell,b) = (43.17^\circ,0.01^\circ)$,
is located on the far-side of the Solar circle.
Its large distance ($\sim$11.4 kpc; Gwinn et al. 1992) 
and low Galactic latitude together
make this object attractive for studying various propagation effects 
due to the intervening medium. The interstellar
scattering in this case is comparable to that in the Vela direction
(Gwinn et al. 1993), but the Vela pulsar is at a much smaller distance ($\le$1 kpc). 

In our efforts to study intrinsic short timescale variability in W3OH 
(Ramachandran et al. 2006; also Laskar et al. 2012)
we were able to estimate and remove possible variability due to 
interstellar scintillation using the fact that the decorrelation bandwidth 
is larger than the measured line widths. Such velocity-resolved
analysis of W3OH data has suggested intrinsic variability on 15-20 minute
time scale (Ramachandran et al. 2006), whereas W49N data show
variations on timescales of 1 hour or longer (Goss et al. 2007: 
talk at IAUS 242, Alice Springs). The timescales of apparent
intrinsic variability, combined with light-travel-time-argument, 
suggest spatial scales for the maser columns to be a few AU 
for the OH sources in W3OH complex (Ramachandran et al. 2006).
The reliability of similar estimation 
in the case of W49N sources depends, however, on how well the contamination
due to interstellar scintillation can be separated from the observed 
variability, particularly when the scintillation decorrelation 
bandwidth may not be much wider than the line widths.
In fact, for the purpose of this variability study,
W49N was included in our VLBA observations as a {\it comparison} source, 
where scattering effects are expected to dominate in the apparent 
relative variability of different spectral features
due to the interstellar scintillations (with decorrelation bandwidth
comparable or smaller than the line widths).
The subsequent analysis
of the data on this source has turned out to be far more fruitful than our
initial expectations, revealing several interesting aspects that
we describe and discuss in this paper.

Measurements of Galactic magnetic fields and of their distribution over 
a range of scales continue to be great interest, given their
important role in the physical processes in the interstellar medium.
Magnetic fields present in molecular clouds become directly measurable 
through the Zeeman splitting of the narrow spectral lines associated 
with several  species (e.g., Sarma et al. 2000, and references therein), 
and those threading across the intervening 
ionized medium manifest themselves through Faraday rotation 
and anisotropic scattering. 
While both the Zeeman splitting of nonmaser lines and the Faraday rotation
depend on the line-of-sight component of the magnetic
field, the anisotropy in scattering relates to the field component in the plane of the sky.  
However, in cases where the Zeeman splitting exceeds the individual line widths
(as is often the case for narrow maser lines such as OH), 
the line separation is then dependent on the {\it total} field strength (Heiles et al. 1993) and not just the component along the line of sight.

In an earlier study of OH maser sources in W49N, 
Desai, Gwinn \& Diamond (1994; hereafter DGD94) 
have found significant anisotropic scattering, attributable to electron-density 
irregularities that are preferentially elongated in the Galactic plane,
due to a magnetic field parallel to the plane.  

Our VLBA observations, and the resultant images of several dozen
maser spots (including a few dozen Zeeman pairs), 
provide significantly improved statistics to investigate
in greater detail the anisotropic scattering, as well as to probe
the distribution of magnetic fields within the W49N complex.  
Some of our findings are in good qualitative agreement with those
of DGD94; they however differ quantitatively from their estimates 
of scatter-broadened
source sizes as well as the mean source position angle (hereafter SPA)
characterizing the apparent shapes
of the maser spots.
These observations provide a rare opportunity to look for another imprint 
of the magnetic fields in the scattering medium, namely, 
possible differential diffraction experienced by opposite circular
polarization components of propagating electromagnetic waves.
This expectation is based on the same refractive index differences
for the two circulars that result in Faraday rotation of 
the position angle of the linear component of incident polarization. 

Our data on 32 Zeeman pairs, providing estimates of magnetic field within
the maser column, enables us to study the distribution and variation
of the magnetic field in this star-forming complex.

In the following section, we describe briefly our observations and data
which form the basis for the analysis and results discussed in the paper. 
Section 3 presents our general findings on the anisotropic scattering observed
in the direction of the OH sources in W49N. The data suggesting significant
differential anisotropic scattering, probed using Zeeman pairs, are 
described in section 4. The structure functions of the 
magnetic field local to the source, as well as the scattering 
parameters estimated for the ensemble, are presented in section 5.
The implications of our results are discussed 
and summarized in the final section.

\section{Our data}
Our visibility data from 12+ hour synthesis observations with 
VLBA (BR107; 2005 July)
were calibrated \& processed in the standard way using AIPS.
The data were self-calibrated; absolute coordinates are not available
due to an absence of phase reference source observations.
The spectral-line image cubes, obtained separately at 1612, 1665 \& 1667 MHz,
provided high angular-resolution images of the set of sources for each
of the transitions, for each of the circular polarizations. 
The total imaged angular extent of about 8 arc-seconds
corresponds to a transverse span of $\sim$~0.5 pc 
at the distance of the source, although a majority of
the sources are within the central 2 arc-second region. 
As a result of exclusion of the outer 2 antennas of 
the VLBA
(St. Croix \& Mauna Kea) from self-calibration, due to lack of 
visibility of the sources on those baselines 
in this imaging, the synthesized beam size across (RA,Dec) is 
$20\times15$ mas. 
(the corresponding spatial resolution is about $\sim$175 AU 
at the distance of W49N).
The lack of visibility on the longest baselines is consistent with,
and is in fact due to,
the apparent angular sizes of the maser sources in the range 20-40 mas.
Each data set consisting of cleaned and restored images across 
240 spectral channels spanning a 22 km s$^{-1}$ 
velocity range, providing velocity resolution of $\sim$0.1 km s$^{-1}$, 
was used to identify discrete maser sources (avoiding cases with
significant velocity gradients). 
 For each of 
clearly identifiable discrete (or isolated) maser sources,
best-fit estimates of 
the {\it deconvolved} source shape and size were obtained, 
in addition to the estimates
of the mean location and velocity, along with estimates of the 
respective uncertainties. For these estimations, we have used a
procedure in AIPS (JMFIT), which facilitates fitting of a 2-d Gaussian
to a source image and returns the best-fit parameters, including errors.
 The shape and size together are parametrized
in terms of major \& minor axes and position angle associated with 
a model ellipse describing the cross-section of a 2-d Gaussian sampled 
at half-power points. Although the absolute position information 
is uncertain by an unknown amount,
the relative positions of the sources at both 1665 \& 1667 MHz 
are reliable within the estimation uncertainties, 
as same self-calibration applies to the data at these two frequencies.
The resultant data on a total of 205 sources 
(194 of these at 1665 \& 1667 MHz) were examined 
for positional proximity of 
left and right circular polarization (LCP and RCP, respectively)
source pairs ($\le$ 10 mas),
and 32 Zeeman pairs were thus identified. 
Figure 1 shows the distribution of all discrete sources
in the RA-Dec plane; the size of the symbol is proportional to
the mean velocity associated with the observed OH line. The circles and
stars denote the sources with RCP and LCP,
respectively. Most of the sources are within an area 
extending from RA-Dec offset (in mas) of (-2000,-2000) to (+4000,+4000).
The velocities (indicated crudely by the symbol size) on the average 
differ systematically between the two halves about the center of this area 
(roughly at RA-Dec offset of 1000,1000 mas). 
An overall correlation between the locations of the sources on the 
sky plane and their line-of-sight velocities, such as that evident for 
the distribution in the Figure 1, is to be expected in system morphologies
dictated by rotation and/or bipolar outflows that are commonly encountered
in star-forming regions (see, for example, Shepherd and Churchwell 1996).

\section{Anisotropic scattering: apparent source sizes and shapes}

Scatter-broadened shapes and orientations of the maser spots provide
valuable information on the nature of the scattering medium.
Figure 2 displays the deconvolved sizes of the sources,
giving the dimensions along the major and the minor axes of the
respective best-fit ellipses. 
The elongations or the axial ratios (ratio of major to minor axis
of the best-fit ellipse) for a majority of the sources are in 
the range 1.5 to 3, indicating
significant anisotropic scattering, consistent with the finding of
DGD94. However, we find the overall apparent
sizes of the sources to be significantly smaller (by a factor of $\ge$ 2)
than those reported by DGD94. Our size estimates are nonetheless
consistent with the suitably scaled values of the scatter-broadening 
of H$_2$O masers reported by Gwinn (1994), 
i.e. about 40 mas (200 $\mu$as times 
[18/1.3]$^2$). Interesting, DGD94 themselves mention about 45 mas as the
expected amount of scatter broadening at 18 cm wavelength for the W49N sources.
Possible reasons for what appears to be an overestimation of the sizes in DGD94
are entirely unclear at present. 
However, we note that the larger apparent angular sizes for the sources 
reported by DGD94 
would necessarily imply significant loss of visibility on a larger fraction 
of the VLBA baselines, an effect which should be readily apparent.

One of the most important aspects discussed by DGD94 
(also see references therein) is that 
the apparent image sizes/shapes are expected to be elongated
orthogonal to the scattering irregularities due to anisotropic 
diffraction.
The presence of a magnetic field would
induce such anisotropy in the electron-density irregularities;
then the implied field direction would be perpendicular to
the resultant image position angle (i.e. SPA).
Based on their limited sample, of the apparent image shapes for 
27 spectral components from 6 of the OH maser sources in W49N,
DGD94 suggested that the apparent 
anisotropy is induced by a magnetic field in and parallel to
the Galactic plane.

In Figure 3, we present a distribution of our estimates of 
the SPAs for our significantly 
($\sim$ 30 times) larger sample.
Some random spread in SPA is apparent in both our
and DGD94's reported SPA values; such a spread ($\sim$5 degrees r.m.s.) 
is not unexpected,
given the estimation uncertainties as well as possible differences 
in the scattering sampled by different lines of sight.
However, we find that our SPA distribution has a significant offset from
the SPA of $\sim$117 degrees, the expected SPA if the density
irregularities are elongated parallel to the Galactic plane.
Our sample gives a mean SPA of 107$\pm$3 degrees, implying 
a significant mean deviation, of $\sim$10 degrees, in the
orientation of the density irregularities from the Galactic plane.
The DGD94 sample
may have been too limited to make such an offset detectable.
We discuss below possible reasons for this systematic offset.
If the large-scale magnetic fields in the intervening region 
has a 10 degree orientation with respect to the Galactic plane, or
if the Galactic warp (e.g., Burton 1988) is responsible for this offset,
then the observed orientations for the scatter-broadened images
would be consistent with a correspondingly revised expectation.
However, W49N is located almost on the farther edge of the inner (Solar)
circle of our Galaxy, at a Galactic longitude of about 43 degrees,
and the effect of the warps is observed to be more pronounced 
in the outer regions of our Galaxy. 
In any case, a look at the magnetic field structure along
the W49N direction would be instructive. 
Interestingly, the sight-line to W49N is through 
the well-known North Polar Spur (NPS) feature (distance to the NPS is
about 250 pc).
From the study, by Wolleben (2007) and others of the NPS,
it is evident that
significantly different magnetic field structure as well as
enhanced scattering would be
expected for the medium within 0.5 kpc of the Sun.
We note that the field orientation in this region, corresponding to the NPS,
might be at a large inclination, if not almost orthogonal,
to the Galactic plane (see Wolleben 2007 for details). 
Also, given that the scattering medium closer to 
the observer is expected to
make relatively higher contribution to the angular broadening 
(e.g., Gwinn et al. 1993, Deshpande and Ramachandran 1998), 
the above mentioned SPA deviation can be caused by the more local scattering
in the NPS region, with density anisotropy at a possibly 
large angle with respect to the Galactic plane. 

The scattering in the W49N direction is unlikely to
be dominated by any single thin screen, but is contributed by the entire 
magneto-ionic medium distributed between the Sun and the source. 
Although a detailed modelling of the magneto-ionic medium incorporating 
the relative strengths and sense of anisotropic scattering, associated
magnetic field strengths and orientations for the different regions 
along the sight-line is needed to assess the net effect, 
we have carried out a simple simulation of propagation through multiple screens
in order to examine the basic situation.
For each screen, a random column-density distribution of free electrons,
following a power-law spatial spectrum (with Kolmogorov index $-11/3$)
was used to simulate a 2-d scattering screen with a mild (1.6) anisotropy
 across a transverse extent of a few times the Fresnel scale 
($a_F = \sqrt{\lambda D}$, where $\lambda$ is the observing wavelength 
and $D$ is the distance to the source).
The orientation of the mean anisotropy in the screen nearest the observer
was chosen to be orthogonal to those in the rest of the screens,
consistent with the possible elongation of density irregularities in NPS region
with respect to that parallel to the Galactic plane.
A model with incoherent combination of the scatter broadening 
from such screens cannot
produce a resultant image with an in-between orientation or SPA, for example, with 10 deg
offset as mentioned above, but would instead dilute 
the aspect ratio.\footnote{ This situation has much in common with 
the polarization state resulting from combination
of signals in two mutually orthogonal (say, linear) polarization states. 
A different orientation is
possible only if the polarized components are at least partially 
mutually coherent, or else, the resulting net
polarization would match that of the dominant of the two components, 
with a reduction in the degree of polarization.}
Hence, the spatial scales of relevant density irregularity in the phase-screens
are likely to be very much larger than the sizes estimated based 
on the observed angular broadening, assuming a single thin screen. 
The relevant spatial scale of density irregularity should be adequately large,
so that emerging wavefront (or field) retains memory of and is the result of 
the {\it coherent} combination of phase structure contributed by 
all sub-screens along the sight-line.
The relevant phase structure should necessarily contain higher order
variations and/or a finite random spread in orientations of phase curvatures.

\section{Differential Anisotropic Scattering: probe with Zeeman pairs}

Thirty two Zeeman pairs can be identified from 
the large sample of sources 
using a positional-proximity criterion  ($\le$ 10 mas).
Based on these Zeeman pairs we can probe a previously unobserved 
aspect of scattering
caused by the magneto-ionic component of the interstellar medium.
A magneto-ionic medium would, in principle, have differing refractive indices 
for signals corresponding to the two hands of 
circular polarization.
Diffractive scintillation and scatter-broadened image shapes should 
therefore differ between the 
LCP and RCP
 due to any line-of-sight component of the magnetic
field in the intervening medium.
Hence scattering-dominated images of even a randomly polarized source might 
show circular polarization in unmatched parts of the images, if 
Faraday rotation is significant.
Macquart \& Melrose (2000) consider this possibility, 
but estimate the effect to be too small (circular fraction 
$\sim 10^{-8}$) to be observable.
Contrary to that expectation, the scatter-broadened images of some of the W49N 
OH maser sources seem to significantly differ in LCP and RCP,
i.e. within a given Zeeman pair.
For the remainder of the Zeeman pairs, the SPA differences are either small 
or within the respective uncertainties.
Figure 4 shows the observed differences in the image shape parameters for
a subset of our sample of Zeeman pairs. The subset includes 
only those cases for which a significant difference ($\ge$ 6$\sigma$) 
in image SPA is observed.
Although there are only a few such Zeeman pairs, 
the SPA differences range between 6 to 30 degrees.
Difference in the line velocities within most of these pairs 
is too small to account for the apparent 
differential scattering.  Position differences are also within a 
few mas (i.e. much smaller compared to the scatter broadening), 
and are unlikely to contribute to the observed SPA differences.
One such example of a significant SPA difference is shown in Figure 5.

To explore the issue further, we have made  preliminary attempts
to simulate signal propagation through a magneto-ionic medium with 
a mild anisotropy.  Here a single, thin scattering screen was simulated,
following the procedure outlined in an earlier section.
For simplicity, a uniform magnetic field is assumed, so that
the phase screens for the two circular polarizations are merely scaled
versions of each other.  The net circular polarization, if
viewed with coarse resolution would be negligible, consistent with
the conclusion of Macquart \& Melrose (2000).  
However, when observed with adequately high angular resolution
(e.g. 10 milli-arc-second), the
simulations do reveal noticeable differential diffractive effects, hence
differing shapes and sizes of scatter-broadened images 
for the two circular polarizations. 
The differential phase contribution due to the Faraday rotation should
be significant, even if relatively small in comparison with the phase
variation common to both polarizations.
The data on rotation measures to several pulsars 
(Manchester et al. 2005; ATNF Pulsar Catalog\footnote{
http://www.atnf.csiro.au/research/pulsar/psrcat})
close to the direction 
of W49N suggest that the expected Faraday rotation at 18 cm would correspond to
about 30 radians of differential phase between the two circular polarization
signals.
In comparison, the phase structure 
function (Armstrong, Rickett \& Spangler 1995) value at 1 AU scale (close to
the Fresnel scale, given the distance to W49N and the wavelength of 18 cm) 
would suggest
an rms phase fluctuation of the order of 300 radians, 
common to both polarizations. Hence, the differential phase contribution
here can lead to noticeable differential scattering.
Of course, more detailed and realistic simulations, i.e. with thick screens, 
are required
to assess these issues in detail. However, the Faraday rotation contribution leading to
differential anisotropic scattering for the two circular polarization 
signals remains
a more natural explanation for the detected differences in source position angles.
If true, the above would be the first such example of this effect in 
an astronomical context, as far as we are aware.

\section{Spatial structure functions}

The sight-lines to the two hundred or so sources  spread across
the W49N region potentially sample variations on a range of transverse scales
within the region, as well as the intervening scattering medium. 
The estimated parameters for this set of scatter-broadened maser sources 
thus provide an excellent opportunity
for probing the spatial scales associated with the variation in these parameters.
However, since this sampling is rather non-uniform,
we prefer to compute spatial structure functions using these data 
instead of trying to estimate the associated spatial power spectra.
Although we treat each of the maser sources as a separate source, 
irrespective of their Zeeman pairing, we compute 
the structure functions separately for the pairs of sources 
with matching and opposite signs of circular 
polarization, along with the structure function for the entire source sample.

Given the limited signal-to-noise ratio in estimation
of most of the parameters, the estimated structure functions 
will have corresponding noise bias. To assess the latter, we also computed
the structure function contribution expected from uncertainties in the
estimations of the relevant parameters.
Figures 6, 7 and 8 show a few examples of the structure functions estimated 
in this manner, where the noise bias is relatively small.
Specifically, in Figure 6 we present the structure function of the 
(line-of-sight) velocity associated with the sources. The clear linear
trend in the log-log plot is consistent with the underlying velocity gradient
across the region, manifesting the possible bipolar outflow morphology.
Figure 7 shows the structure functions for the scatter-broadened 
size of the sources, in this case the minor axis associated with the
elliptical source shape. 
Significant variations in the scattering property is apparent
on an angular scale of about 200-300 mas.
For an average distance to the scatterer of about half-way to the
source (5.7 kpc), 
this angular scale would correspond to a spatial scale of 1000-1500 AU
over which scatter broadening appears to change
significantly. 
 
The spatial structure function computed for the magnetic field using estimates 
available for the set of Zeeman pairs, as presented in Figure 8,
shows no significant monotonic trend. This poor sampling of the 
structure function is largely due to the 
uneven and sparse sampling of the spatial 
scales that the available Zeeman pairs in this region offer. 
However, the scale on which the magnetic field appears to be decorrelated
is about 150 mas (or 1500 AU spatial scale at the source distance).
For 10 out of the 32 pairs, the Zeeman splitting is smaller than the line widths.
Hence in these cases, the estimated magnetic field would correspond to only
the line of sight component.

\section{Summary}

The OH maser sources in W49N show anisotropic scattering; 
the apparent scatter broadening is much less than reported earlier.
The implied scintillation decorrelation bandwidth, 
corresponding to the observed sizes
of the W49N sources ($\ge$20 mas), would be about 100 Hz 
(typical line widths are 2.5 kHz).
This would correspond to a velocity width of 20 m/s, less than 
our velocity resolution (100 m/s). The sizes reported by DGD94 
would have implied even narrower decorrelation bandwidths.
Hence, we expect interstellar scintillations to be significantly quenched,
reducing potentially contaminating contribution to any apparent intensity
variability.

The position angles of the source images deviate significantly from the value 
expected if scattering density irregularities were to be ``stretched" 
due to magnetic field strictly aligned parallel to the Galactic plane,
indicating significant scatter-broadening contribution from
differently aligned density irregularities. This contribution may 
be associated with the North Polar Spur, within a distance of 0.5 kpc. 
Some of our data also reveal differential scattering during propagation 
of the two circular polarizations, likely caused by Faraday rotation.
Through various aspects discussed above, the attractive, but yet unexplored, 
potential of the high-resolution maser observations for 
probing the intervening magneto-ionic medium is certainly evident. 

The structure function analysis of the various parameters suggests
spatial scales of a few thousand astronomical units over which
the scattering properties in the intervening medium as well as
the magnetic field in the maser region appear to be changing.  

\section*{Acknowledgements:} It is a pleasure to acknowledge the contributions 
from R. Ramachandran and Sarah Streb at different stages of the reported 
work. We thank Don Melrose and the anonymous referee for their valuable comments. 
AAD gratefully acknowledges fruitful discussions with Rajaram Nityananda,
Simon Ellingsen and John Dickey. 
The National Radio Astronomy Observatory is a facility of 
the National Science Foundation operated under a cooperative agreement 
by Associated Universities, Inc.\\

{\it Facilities:} \facility{VLBA}

\clearpage

\begin{figure}[b]
 \vspace*{-0.25 cm}
\begin{center}
\includegraphics[width=5.0in,angle=-90]{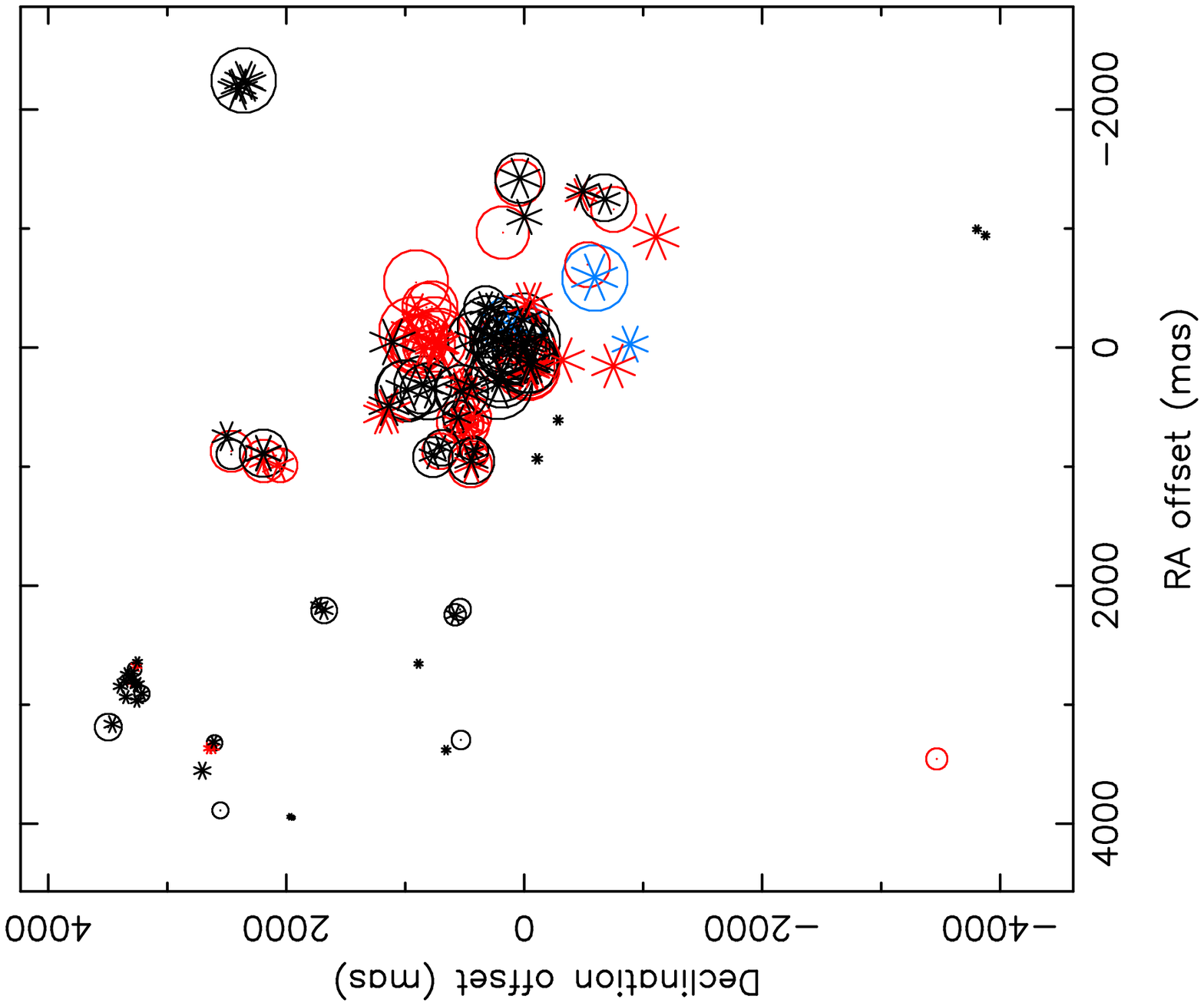} 
 \caption{Observed distribution of W49N OH maser sources in RA-Dec. 
2000 mas corresponds to 0.11 pc at the 11.4 kpc distance of W49N.
The symbols in blue, red and black denote the sources at 1612, 1665 and 1667 MHz, respectively. The circles and stars refer to the Left and the Right hand circular polarization, respectively.
The symbol sizes 
are scaled proportional to the velocity 
(which are in the range +2 to +21 km s$^{-1}$). 
The velocity-position correlation 
and an implied bipolar nature is apparent from the majority of 
the sources in the sample. (See the main text for details.)
}
   \label{fig1}
\end{center}
\end{figure}

\begin{figure}[b]
 \vspace*{-0.25 cm}
\begin{center}
 \includegraphics[width=5.0in,angle=-90]{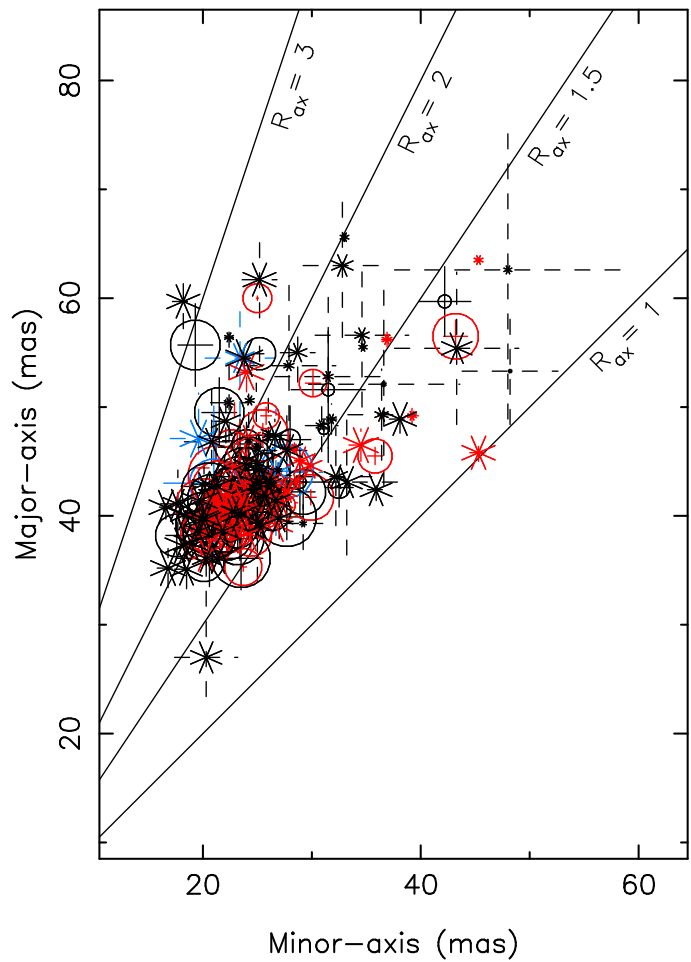} 
 \caption{Scatter plot of the Major axis versus the Minor axis 
values estimated for the 205 spots, along with their 
1$\sigma$ uncertainties
indicated by the error bars. 
The correspondence of the symbols, including their colour and sizes, with the
source parameters is
same as that in the Figure 1.
The four lines across the plot (from bottom to top) correspond to
axial ratio, R$_{ax}$, of 1, 1.5, 2. and 3, respectively.
A majority of the sources have axial ratios in the range 1.5 to 2.}
   \label{fig2}
\end{center}
\end{figure}

\begin{figure}[b]
 \vspace*{-0.25 cm}
\begin{center}
 \includegraphics[width=4.0in,angle=-90]{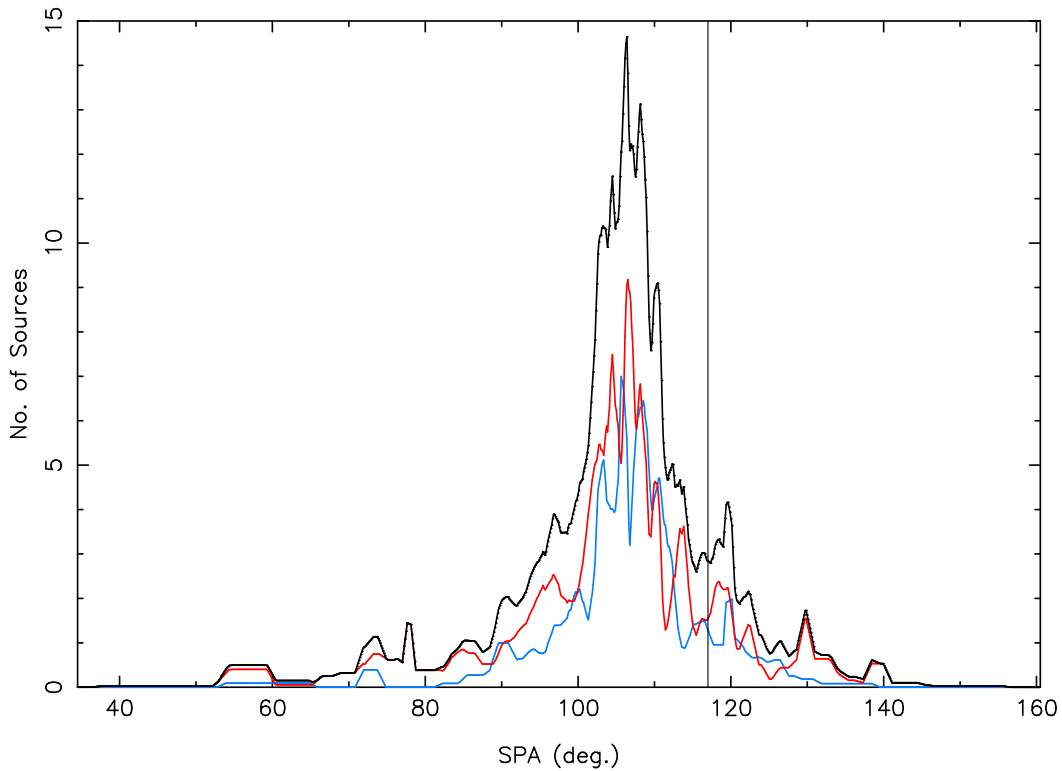} 
 \caption{Distribution of the source position angles (SPAs) of the 
scatter-broadened images 
for our full set of OH maser sources in W49N is shown (black profile).
The vertical line
at 117 degrees, shown for reference, corresponds to
an angle orthogonal to the Galactic plane. 
This is the value of SPA that would be 
expected from the anisotropic scattering 
due to electron density irregularities, if these 
are elongated parallel to the Galactic plane. 
The distributions of SPA shown here are after accounting 
for uncertainties in individual measurements 
(the blue and red profiles correspond to RCP and LCP, respectively).
This distribution is used to estimate the mean SPA of 
the observed scatter-broadened images (107$\pm$3 degrees).}
   \label{fig3}
\end{center}
\end{figure}

\begin{figure}[b]
 \vspace*{-0.25 cm}
\begin{center}
\includegraphics[width=4.7in,angle=-90]{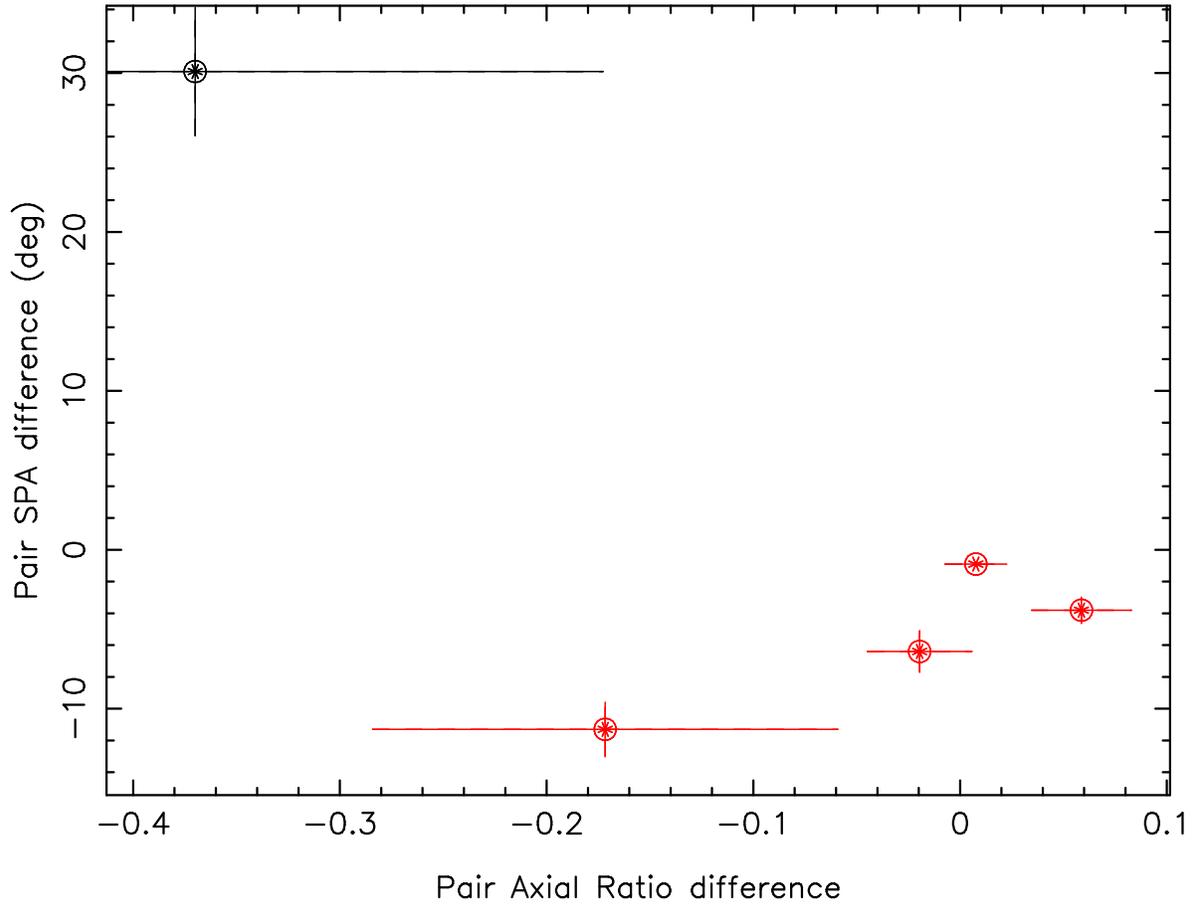} 
 \caption{The SPA differences within Zeeman pairs versus the 
respective differences in axial ratios
for the cases where significant SPA differences are apparent. 
The error bars indicate $\pm1\sigma$ uncertainties.
Only one of these four sources is at 1667 MHz (in black),
while the other three (in red) are at 1665 MHz.
The axial ratios are consistent with equality of
RCP and LCP}.
   \label{fig4}
\end{center}
\end{figure}

\begin{figure}[b]
 \vspace*{-0.25 cm}
\begin{center}
\includegraphics[width=6.5in,angle=0]{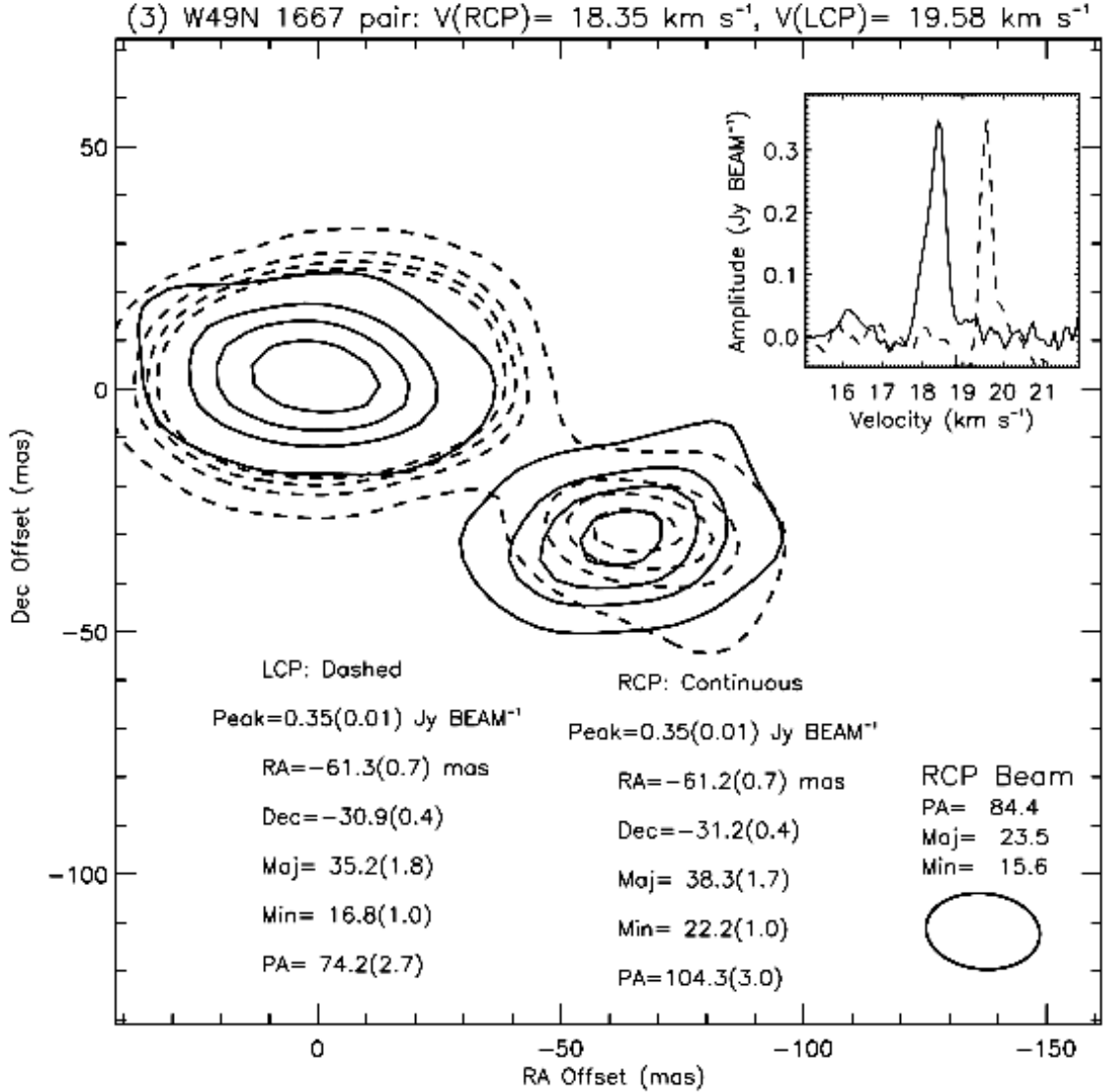} 
 \caption{Apparent images are shown for the sources (at RA-DEC offsets -60 and -30 mas respectively)
in both hands of circular polarization, i.e.
for the Zeeman pair, where a significant difference (about 30 degrees) in the position angles of the 
scatter broadened shapes in the two circular polarization is evident.
The associated line profiles are shown in the upper inset. 
The source at  RA-DEC offsets of about 0,0,
is also shown for comparison, where the image ellipses 
in the two circular polarizations have
mutually consistent orientations.}
   \label{fig5}
\end{center}
\end{figure}

\begin{figure}[b]
 \vspace*{-0.25 cm}
\begin{center}
\includegraphics[width=5.0in,angle=-90]{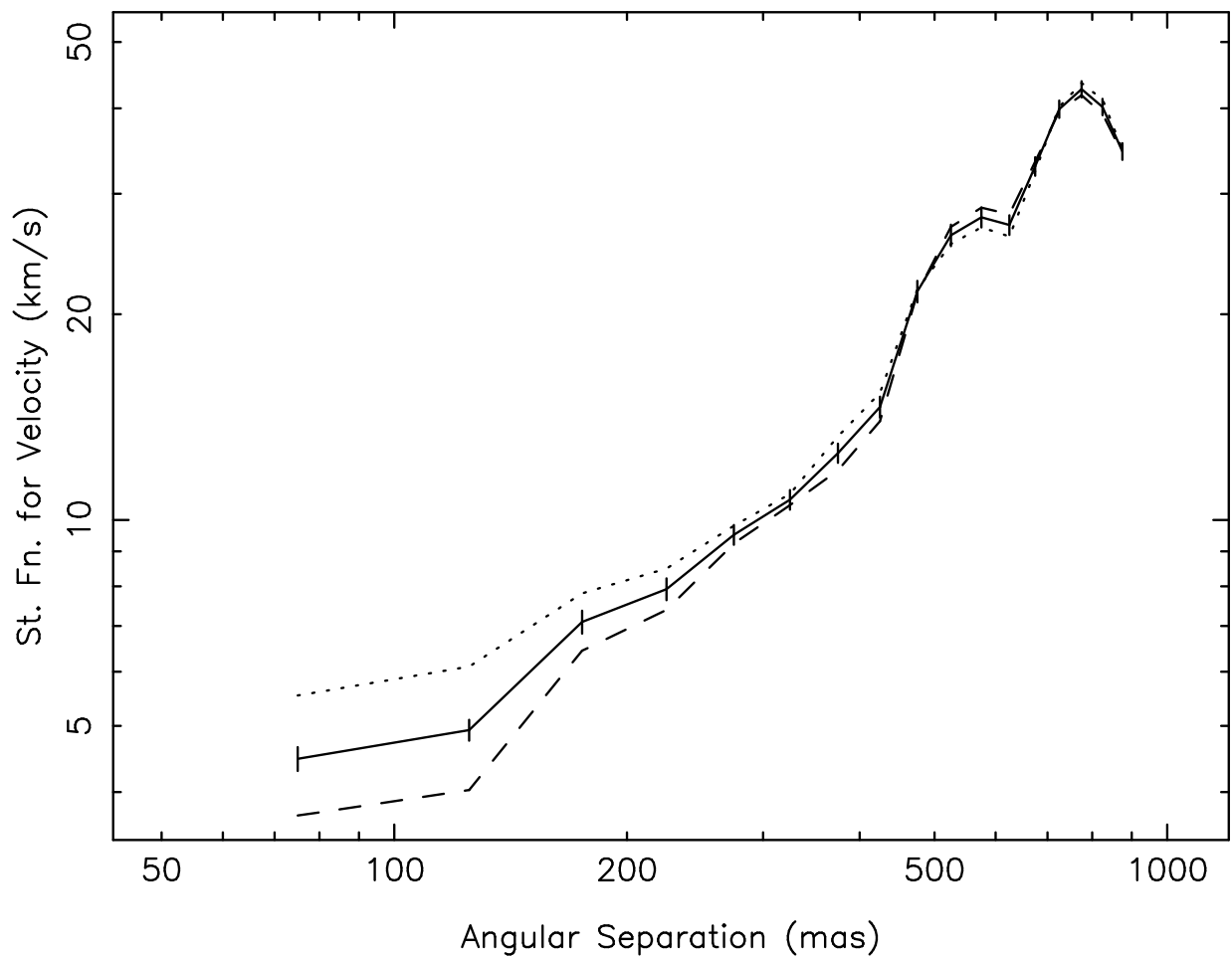} 
 \caption{Spatial structure function of the line-of-sight velocity. 
The structure function values are in (km s$^{-1}$)$^2$.
The solid-line curve corresponds to the contribution from all source pairs, while the
curves drawn with dashes and dots correspond to source pairs with matching 
and opposite circular polarization, respectively. The observed trend in 
the structure function is consistent with the velocity gradient 
across the source. Such position-velocity correlation is typical of 
a systematic velocity field, such as those found in bipolar outflows.
The 1$\sigma$ error bars (shown only for the solid-line 
curve) include contributions from both the
measurement and the statistical uncertainties in the estimates.}
   \label{fig6}
\end{center}
\end{figure}

\begin{figure}[b]
 \vspace*{-0.25 cm}
\begin{center}
\includegraphics[width=5.0in,angle=-90]{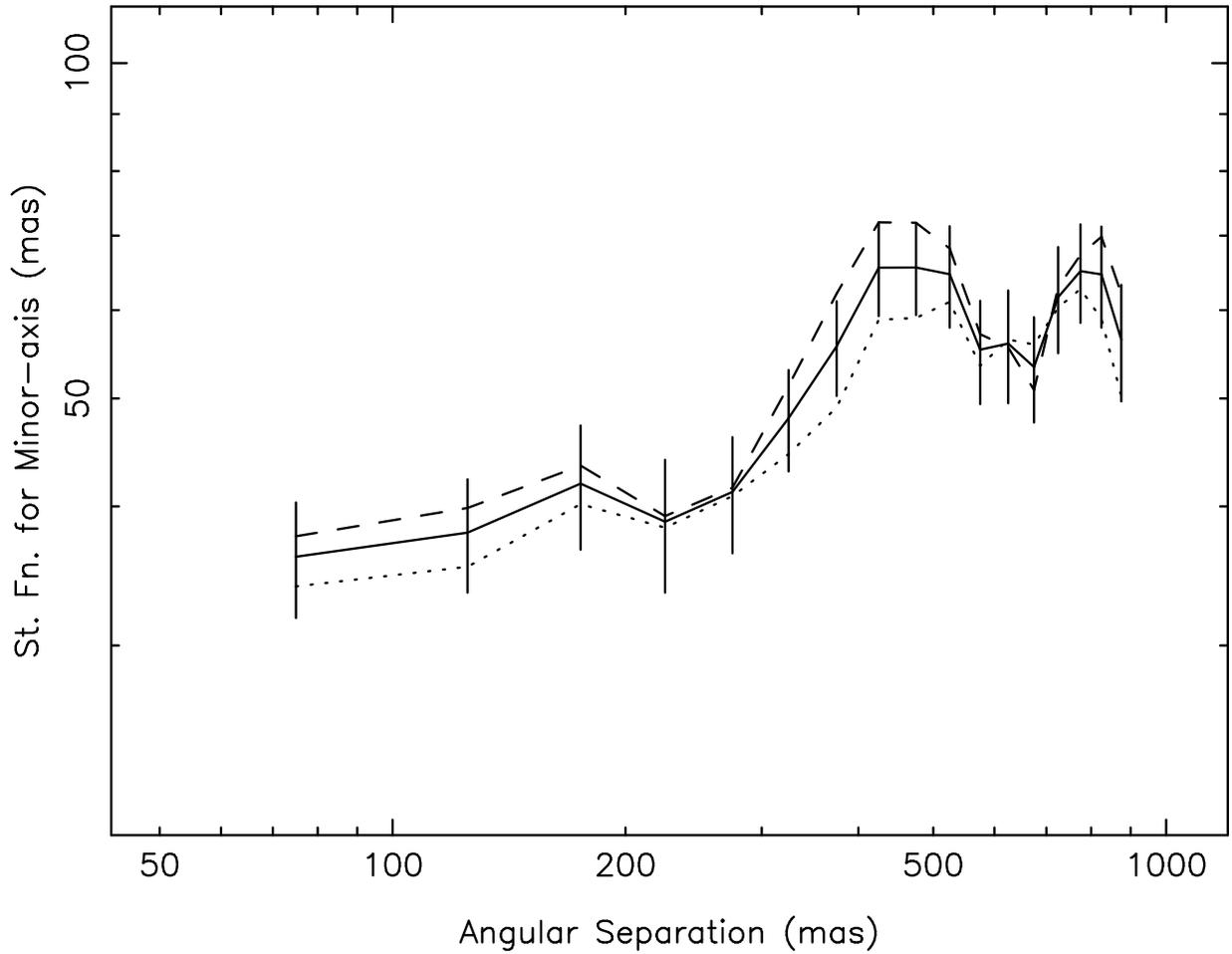} 
 \caption{Similar to Figure 6, for the minor-axis size associated with 
the scatter-broadened images. The structure function is in units 
of (mas)$^2$. The average trend in the above figure suggests 
that scattering properties have significant 
differences on angular scales larger than 200 mas.}
   \label{fig7}
\end{center}
\end{figure}

\begin{figure}[b]
 \vspace*{-0.25 cm}
\begin{center}
 \includegraphics[width=5.0in,angle=-90]{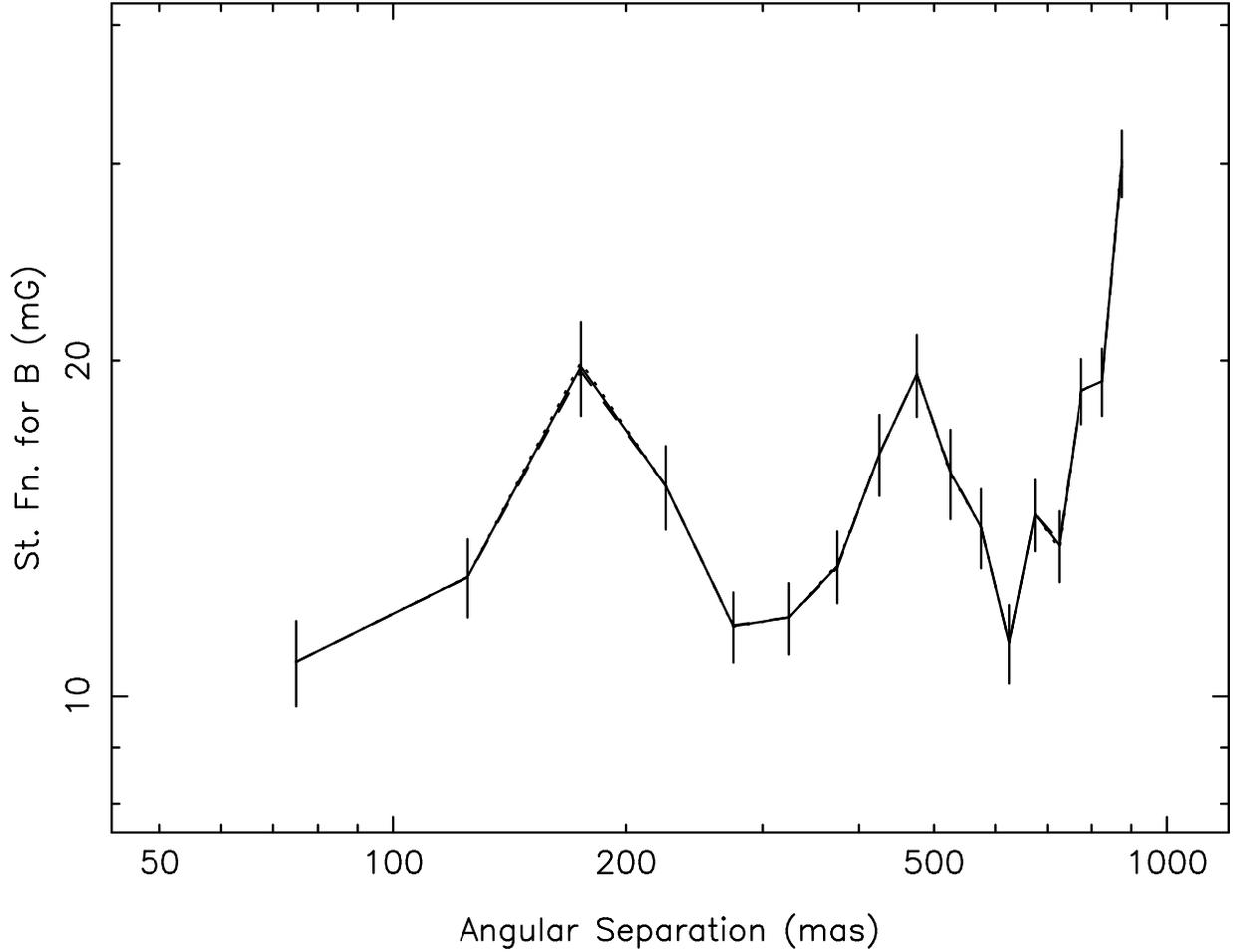} 
 \caption{Similar to Figure 6, for the magnetic field estimated for the Zeeman pairs.  The structure function is in units of (mG)$^2$.
The strength of the magnetic field within the W49N region 
seems to show variations on angular scales
of 100 mas and beyond. Owing to poor statistics, this structure function 
estimate is sensitive to binning across the angular separation.
See main text for details.}
   \label{fig8}
\end{center}
\end{figure}

\end{document}